\documentclass[12pt]{iopart}
\usepackage{amssymb}
\usepackage{iopams,bm,verbatim}
\usepackage[T1]{fontenc}
\usepackage{amsthm}

\def\Journal#1#2#3#4{{#4} {\it #1} {\bf #2}, #3 }

\def\ud{\textrm{d}}

\def\bmp{\bm{\partial}}

\def\FE{\mathcal{Y}}
\def\E{\mathcal{I}}
\def\H{\mathcal{J}}
\newtheorem*{thm}{Theorem}

\newcommand{\df}{{\mbox{\rm d}}}
\newcommand{\smfrac}[2]{{\textstyle{#1\over#2}}}
\def\half{\smfrac{1}{2}}

\begin{document}

\title{On the relation between the Einstein field equations and the Jacobi-Ricci-Bianchi system.}

\author{N.~Van den Bergh}

\address{Ghent University, Department of Mathematical Analysis EA16, \\ Galglaan 2, 9000 Ghent, Belgium}
\eads{\mailto{norbert.vandenbergh@ugent.be}}

\begin{abstract}
The 1+3 covariant equations, embedded in an extended tetrad formalism and describing a spacetime with an arbitrary energy-momentum distribution, are reconsidered. 
It is shown that, provided the  1+3 splitting is performed with respect to  a generic time-like congruence with tangent vector $\bm{u}$, the Einstein field equations
can be regarded as the integrability conditions for the Jacobi and Bianchi equations together with the Ricci equations for $\bm{u}$. 
The same conclusion holds for a generic null congruence in the Newman-Penrose framework.
\end{abstract}


\section{Introduction}

Although the Einstein field equations\footnote{As our discussion applies to a completely \emph{general} energy-momentum tensor, a possibly non-zero cosmological constant has been 
absorbed in the latter.} 
\begin{equation}\label{EFE}
R_{ij}-\frac{1}{2}R g_{ij} = T_{ij} 
\end{equation}
have a deceptively simple appearance, when written out in terms of coordinates and the components of the metric, they form an unwieldy set of second order non-linear partial differential 
equations. This is one of the reasons why, during the past half century and 
in several areas of gravity research (explicit construction and classification of exact solutions \cite{Karlhede, Kramer}, cosmological perturbations and gravitational waves
\cite{Hawking66,DunsbyBassEll,EllisBruni89,Tsagas08}, numerical relativity \cite{JanCarBin92,Friedrich96,vanPuttenEardley96,vanElstUggla97,ShinYon00}, 
fundamental aspects \cite{CahenDebeverDefrise67,Deser76,Pleb77,Ash86,Capovilla1990}), focus has been shifting towards tetrad formulations of the theory.
In this paper I consider the so called covariant 1+3 formalism and the related orthonormal tetrad formalism, which continue to play an important role in the context of cosmology, as well as the Newman-Penrose
formalism.
The structure of the governing equations of relativistic cosmology, including their consistency, in both formalisms has by now been discussed in a large number of 
papers\cite{Ell67, MacCallum_Cargese, Maartens2, MaaLesEll97, Els96, Vel97, Uggla_etal2003, WainwrightEllis} (for a recent review see the book by Ellis, Maartens and MacCallum \cite{EMM}),
while integrability of the equations in general tetrad formalisms has been discussed in detail by Edgar\cite{Edg80,Edg92} and MacCallum\cite{Mac_integrability}, following earlier work by 
Papapetrou\cite{Pap70,Pap71,Pap71a}. There still remain some question marks regarding the redundancies present in these formalisms: while it is obvious that there are a number of algebraic 
interdependences, the issue of differential relations (more particularly the possibility of obtaining certain equations as integrability conditions of others) is less clear.
This paper aims to have a closer look at these differential interdependences among the sets of Jacobi, Ricci and Bianchi equations on the one hand and the field equations on the other.
As the commutator relations form a natural ingredient of such an investigation, the Jacobi equations will play a prominent role:
I first repeat some basic facts about tetrad formalisms in section 2, closely following paper \cite{Mac_integrability} and pointing out some subtleties with 
regard to the Jacobi equations.  In section 3 I show that the field equations turn out to be the integrability conditions of the Jacobi-Ricci-Bianchi set, provided the time-like congruence, with respect 
to which the 1+3 splitting is carried out, has a non-vanishing acceleration which is not an eigenvector of $\sigma_{ab}+\omega_{ab}$. In section 4 I show that the same conclusion holds for a null-congruence $\bm{k}$, 
provided that in a Newman-Penrose tetrad $(\bm{m}, \overline{\bm{m}},\bm{\ell},\bm{k})$ the null-congruence $\bm{k}$ satisfies $- \kappa \equiv k_{a;b}k^b m^a \neq 0$.

\section{Tetrad formalisms} 

As usual spacetime is represented by a pseudo-Riemannian manifold and all calculations apply to an open set of spacetime. Notations and conventions are as in \cite{Kramer}. 
A 1+3 splitting is accomplished locally by choosing 
a congruence of time-like worldlines with unit tangent vector field $\bm{u}$. This congruence may be interpreted as the 4-velocity field of a family of observers, but no further 
restrictions apply: the congruence not necessarily consists of trajectories orthogonal to some family of space-like hypersurfaces, neither needs the unit tangent vector field to be 
parallelly transported or be invariantly 
defined (although in most applications one of these situations might occur). A set of (smooth) basis vector fields is then constructed by erecting at each point a spatial triad $\partial_\alpha$ 
orthogonal to $\partial_0 = \mathbf{u}$ (note that $\partial_a$ are not partial derivatives, but frame derivatives\footnote{Indices a,b,c, \ldots and i,j,k, \ldots are respectively frame indices and 
coordinate indices taking the values 0,1,2,3; greek indices $\alpha, \beta, \gamma, \ldots$ are frame indices taking the values 1,2,3.}: $\partial_a = {e_a}^i \partial / \partial x^i$).
Denoting the dual basis of one-forms as $\bm{\omega}^a={\omega^a}_i \ud x^i$ and the Levi-Civita connection as $\nabla$, the connection coefficients
and connection one-forms are given by $\nabla_{\partial_a} \partial_b= \Gamma^c_{~ba}\partial_{c}$ and $\bm{\Gamma}^a{}_b={\Gamma}^a{}_{bc}\bm{\omega}^c$ respectively.
The Cartan structure equations read then
\begin{equation}
\df \bm{\omega}^a=-\bm{\Gamma}^a{}_b\wedge \bm{\omega}^b \label{cartan1}
\end{equation}
and
\begin{equation}
\df \bm{\Gamma}^a{}_b+\bm{\Gamma}^a{}_c\wedge \bm{\Gamma}%
^c{}_b=\bm{R}^a{}_b \label{cartan2}~,
\end{equation}
$\bm{R}^a{}_b$ being the curvature 2-forms. 
The integrability conditions for (\ref{cartan1},\ref{cartan2}) are given by the first and
second Bianchi identities, $\df^2 \bm{\omega}^a = 0$ and $\df^2 \bm{\Gamma}^a{}_b =0$, or  
\begin{eqnarray}
& \bm{R}^a{}_c \wedge \bm{\omega}^c = 0~,\label{bianchi1}\\
& \df \bm{R}^a{}_b - \bm{R}^a{}_c \wedge \bm{\Gamma}^c{}_b + \bm{\Gamma}^a{}_c \wedge \bm{R}^c{}_b = 0~. \label{bianchi2}
\end{eqnarray}

Henceforth it will be assumed, as is usually done in tetrad formulations of general relativity, that the basis is \emph{rigid}, in the sense that the metric components $g_{ab}$ are constants,
\begin{equation}
\df g_{ab} = \df (\partial_a . \partial_b) = 0~, \label{metricity}
\end{equation}
with, for an orthonormal tetrad, $g_{ab}=\textrm{diag}(-1,\,+1,\,+1,\,+1)$.
Raising and lowering tetrad indices with $g_{ab}$ and its inverse and defining $\Gamma_{ab}=g_{ac}{\Gamma^c}_b$ (\ref{metricity}) implies $\bm{\Gamma}_{ab}=\bm{\Gamma}_{[ab]}$ or
\begin{equation}\label{Gammasymm}
 \Gamma_{(ab)c}=0~,
\end{equation}
such that (\ref{cartan1}) completely defines the connection.

The commutation coefficients ${{\gamma}^c}_{ab}$ are defined by
\begin{equation}
[\partial_a, \partial_b]={{\gamma}^c}_{ab}\partial_{c}~, \label{commut}
\end{equation}
or
\begin{equation}
             {{\gamma}^c}_{ab} = 2 {\omega^c}_j {e_{[b}}^j{}_{,a]}~, \label{riccirot}
\end{equation}
a comma denoting a partial derivative 
 or a frame derivative, depending on the index used.
The first Cartan equations (\ref{cartan1}) express that the connection is torsion-free and hence relate the connection and commutation coefficients by
\begin{equation}\label{commdef}
 {{\gamma}^c}_{ab}=2{\Gamma}^{c}{}_{[ba]}~,
\end{equation}
which, for a rigid basis, by (\ref{Gammasymm}) is equivalent with
\begin{equation}\label{commdef2}
\Gamma_{cab}= \frac{1}{2}({\gamma}_{bca}+{\gamma}_{acb}-{\gamma}_{cab})~.
\end{equation}
Introducing the components $R^a{}_{bcd}$ of the curvature two-forms by
\begin{equation}
\bm{R}^a{}_b=\half R^a{}_{bcd}\bm{\omega}^c \wedge  \bm{\omega}^d~,
\end{equation}
the second Cartan equation becomes
\begin{equation}\label{cartan2bis}
 {R^a}_{bcd}={\Gamma ^a}_{bd,c}-{\Gamma ^a}_{bc,d}+{\Gamma ^e}_{bd}
{\Gamma ^a}_{ec}
-{\Gamma ^e}_{bc} {\Gamma ^a}_{ed} - {{\gamma}^e}_{cd} {\Gamma ^a}_{be}~.
\end{equation}
The first Bianchi identity (\ref{bianchi1}), which can also be written as
\begin{equation}\label{bianchi1bis}
 R_{a[bcd]} = 0~,
\end{equation}
is then seen to be equivalent with the Jacobi identity for a triple of basis vectors $\partial_a$,
\begin{equation}
\left[\partial_{[a},\left[\partial_b,\partial_{c]}\right]\right]=0~, \label{Jacobi}
\end{equation}
or
\begin{equation}\label{Jacobibis}
\partial_{[a}{{\gamma}^d}_{bc]}-{{\gamma}^d}_{e[a} {{\gamma}^e}_{bc]} =0~,
\end{equation}
while the second Bianchi identity (\ref{bianchi2}) can be written as (a semi-colon denoting the covariant derivative with respect to ${\nabla}$)
\begin{equation}\label{bianchi2bis}
 R^a{}_{b[cd;e]} = 0~.
\end{equation}
In a (pseudo-)Riemannian space the Riemann tensor also must satisfy
\begin{equation} 
R_{abcd}=-R_{bacd}=-R_{abdc}=R_{cdab}~. \label{Riesymm}
\end{equation}
There is some confusion about whether or not these symmetry conditions have to be imposed as extra conditions\cite{Mac_integrability}: if the connection is defined in
terms of the commutator coefficients by (\ref{commdef2}) and if the Riemann tensor is defined by (\ref{cartan2bis}), then $R_{abcd}=g_{ae}{R^e}_{bcd}$ is automatically anti-symmetric in the first and second pair 
of indices. Provided that the Jacobi equations (\ref{Jacobibis}) hold, this implies also $R_{abcd}=R_{cdab}$. 
This is what happens in a so-called 
\emph{minimal tetrad formulation}\cite{Edg80, EMM}\footnote{The versions in the cited works differ slightly.}, where one regards the tetrad components 
${e_a}^i$, the ${\gamma^c}_{ab}$ and the matter fields as functions 
to be solved for, with the governing equations being the component forms (\ref{riccirot}) and (\ref{Jacobibis}) of respectively the first Cartan and Jacobi equations and the field equations (\ref{EFE}) together
with the matter field equations.
Note the crucial role of the Jacobi equations, which, via the symmetry conditions (\ref{Riesymm}), 
also imply that the Ricci tensor, defined by contraction of (\ref{cartan2bis}),
\begin{equation}
R_{bd}={R^c}_{bcd}=i_{{\partial}_b}i_{{\partial}_m} {\bm{R} ^m}_d~,
\end{equation}
is symmetric. In addition the Einstein tensor is then divergence-free, such that the matter field equations must be compatible with ${T^{ab}}_{;b}=0$. 
In this approach the remaining trace-free part of the curvature is given by the Weyl-tensor, which is \emph{defined} by 
\begin{equation}\label{riemanndecomp}
 C_{abcd}=R_{abcd}- (g_{a[c} R_{d]b} +g_{b[d} R_{c]a} )+\smfrac{1}{3} R g_{a[c} g_{d]b}~.
\end{equation}
As noticed in \cite{Edg80} one can remove the tetrad components ${e_a}^i$ from the system and view the operators $\partial_a$ as the third set of variables alongside the ${\gamma^c}_{ab}$ and the matter fields.
The governing equations are then the commutator relations (\ref{commut}), formally describing the behaviour of the $\partial_a$ `variables', the
Jacobi equations (\ref{Jacobibis}) and the field equations (\ref{EFE}) together with the matter field equations.\\

Because of the special role played by the Weyl tensor in the construction of exact solutions and in the classification of spacetime geometries, it has become customary to include the 
components of the Weyl tensor as extra variables in the previous system. In the case of an orthonormal tetrad formalism one decomposes then the Weyl tensor 
with respect to a time-like congruence defined by $\bm{u}=\partial_0$ into its electric and magnetic 
components,\footnote{The Levi-Civita tensor $\bm{\eta}$
is normalized such that (tetrad indices!) $\eta_{0123}=-1$; we also define $\varepsilon_{abc}=\eta_{abcd}u^d$, while $h_{ab}$ is the projector in the instantaneous 3-spaces orthogonal 
to $\bm{u}$, defined by $h_{ab}=g_{ab}+u_au_b$.}
\begin{equation}\label{weyldecomp}
{C_{ab}}^{cd}=4(u_{[a}u^{[c}+h_{[a}^{~~[c})E_{b]}^{~~d]}+2\varepsilon_{abe}u^{[c
}H^{d]e}+2\varepsilon^{cde}u_{[a}H_{b]e}~,
\end{equation}
or
\begin{eqnarray}
 E_{ab}=C_{acbd}u^c u^d~,\\
 H_{ab}=\frac{1}{2}{\eta_{ac}}^{ef}C_{efbd}u^c u^d~.
\end{eqnarray}
Provided the Jacobi equations hold and the Riemann tensor is defined by (\ref{cartan2bis}), both $E_{ab}$ and $H_{ab}$ are then symmetric and trace-free.\\

Alternatively one can set up an \emph{extended tetrad formalism}\cite{EMM}, in which the fundamental variables are the $\partial_a$ operators, the ${\gamma^c}_{ab}$ coefficients and the symmetric tensors $E_{ab}$, $H_{ab}$, 
$T_{ab}$ ($E_{ab}$ and $H_{ab}$ being also trace-free) and the matter fields. It is furthermore customary to 
split $T_{ab}$ as
\begin{equation}\label{Tdecomp}
 T_{ab}= \rho u_a u_b + p h_{ab} +2 q_{(a}u_{b)}+\pi_{ab}~,
\end{equation}
where the energy current density vector $q_a$ and the anisotropic pressure tensor $\pi_{ab}=\pi_{(ab)}$ are orthogonal to $u_a$ and $\pi_a^a=0$. 
In this extended formalism the Riemann tensor \emph{is defined by} (\ref{riemanndecomp}), using (\ref{EFE}, \ref{weyldecomp}) to express it in terms of the variables $E_{ab}$, $H_{ab}$, $\pi_{ab}$, $q_{a}$, $\rho$ 
and $p$\footnote{For example: $R_{1214}= H_{31}-\half q_2$, $R_{1212}=\smfrac{1}{3} \rho -\half \pi_{33}-E_{33}$, \ldots}. The symmetry and trace properties of these variables guarantee the 
conditions (\ref{bianchi1bis}, \ref{Riesymm}) and the governing equations become  
the second Bianchi equations (\ref{bianchi2bis}) 
(written in terms of $E_{ab}$, $H_{ab}$, $\pi_{ab}$, $q_{a}$,$\rho$, $p$ and their covariant derivatives)
and (\ref{cartan2bis}), implying (\ref{Jacobibis}). Note that (\ref{cartan2bis})
is now a set of partial differential equations for the ${\gamma^a}_{bc}$ and \emph{not} a definition for the Weyl tensor and that its contraction automatically yields the Einstein field equations 
(\ref{EFE}). An equivalent approach is to implement (\ref{cartan2bis}) by requiring that the Ricci equation holds, when applied to \emph{all four} of the basis vector 
fields:
\begin{equation}\label{Ricciid}
 {w^a}_{;c;d}-{w^a}_{;d;c}= {R^a}_{bdc} w^b~,
\end{equation}
again with appropriate substitutions for the Riemann tensor in terms of $E_{ab}$, $H_{ab}$ and $T_{ab}$. 

A third approach to implement (\ref{cartan2bis}) (which is the one usually followed in cosmological applications, see for example \cite{vanElstUggla97, Uggla_etal2003} or \S 6.6 of \cite{EMM}) 
is to impose \emph{only} the Ricci 
equation applied to $\bm{u}=\partial_0$,
\begin{equation}\label{Ricci0}
 {u^a}_{;c;d}-{u^a}_{;d;c}= {R^a}_{0dc}
\end{equation}
and to add the remaining field equations (\ref{EFE}), together with the  
Bianchi equations (\ref{bianchi2bis}) and the Jacobi equations (\ref{Jacobibis}) (which now \emph{not} automatically hold, as only part of the Ricci equations is used).
The variables of the resulting system, which obviously contains a number of redundancies, can be written explicitly\footnote{As explained above also the $\partial_a$ operators themselves
should be considered as variables of the system, with the commutator
relations viewed as their governing equations.} as $\dot{u}_\alpha$, $\omega_\alpha$, $\theta$, $\sigma_{\alpha \beta}$, $a_{\alpha}$, 
$n_{\alpha \beta}$, $\Omega_\alpha$, $\rho$, $p$, $q_{\alpha}$, $\pi_{\alpha \beta}$, $E_{\alpha \beta}$, $H_{\alpha \beta}$, 
with the kinematical quantities $\dot{u}_\alpha$, $\omega_\alpha=\frac{1}{2}\varepsilon_{\alpha \beta \gamma} 
\omega^{\beta \gamma}$, $\theta$ and $\sigma_{\alpha \beta}$ being defined in the usual 
way by splitting $u_{a;b}$ as
\begin{equation}\label{u_decomp}
  u_{a;b}=-\dot{u}_{a}u_{b}+\sigma_{ab}+\smfrac{1}{3}\theta h_{ab}+\omega_{ab}~, 
 \end{equation}
with $\sigma_{ab}=\sigma_{(ab)}$, $\sigma^a{}_a = 0$, $\omega_{ab}=\omega_{[ab]}$, $\sigma_{ab} u^b =\omega_{ab} u^b = \dot{u}_{a}u^a =0$.

\noindent
$\Omega^a=\frac{1}{2}\eta^{abcd}u_b\partial_c\cdot \dot{\partial}_d$ is the local angular velocity, in the rest-frame of an observer
with four-velocity $u$, of the triad $\partial_{\alpha}$  with respect to a set of Fermi-propagated axes (using MacCallum's convention\cite{MacCallum_Cargese}) and
$n_{\alpha \beta}$, $a_\alpha$ are the Kundt-Sch\"ucking-Behr variables\cite{EllisMac} parametrizing the purely spatial 
commutation coefficients ${{\gamma}^\alpha}_{\beta\gamma}$. 
The commutation coefficients can be read off from
\begin{eqnarray}\label{comm_explicit}
{}[\partial_0,\partial_\alpha] &=& \dot{u}^\alpha \partial_0 - \left(\smfrac{1}{3} \theta\delta_\alpha^{\beta}+\sigma_\alpha^{\beta}
+{\varepsilon^\beta}_{\alpha\gamma}(\omega^\gamma+\Omega^\gamma)\right) \partial_\beta~, \\
{}[\partial_\alpha, \partial_\beta ] &=&  -2\varepsilon_{\alpha\beta\gamma}\omega^\gamma\partial_0 + \left(2a_{[\alpha}\delta^\gamma_{\beta]}+
\varepsilon_{\alpha\beta\delta} n^{\delta\gamma}\right)\partial_\gamma
\end{eqnarray}   
and the connection one-forms read accordingly\footnote{For readability multiplets of equations, which can be obtained from each other by cyclic permutation of the spatial indices, 
will henceforth be represented by a single equation.} 
\begin{eqnarray}
\bm{\Gamma}_{10} = \dot{u}_1 \bm{\omega}^0 +\theta_1 \bm{\omega}^1+(\sigma_{12}+\omega_3)\bm{\omega}^2+(\sigma_{13}-\omega_2)\bm{\omega}^3~,\\
\bm{\Gamma}_{12} = -\Omega_3 \bm{\omega}^0 + (n_{13}-a_2) \bm{\omega}^1+(n_{23}+a_1)\bm{\omega}^2+\half(n_{33}-n_{11}-n_{22})\bm{\omega}^3.
\end{eqnarray}
In cosmological applications it is furthermore customary to define a fully projected covariant derivative $\tilde{\nabla}$, having the property that for an arbitrary tensor $S$
\begin{equation}
\tilde{\nabla}_{a}S^{c\ldots d}{}_{e\ldots f} = h_a{}^b h^c{}_p\ldots h^d{}_q
h_e{}^r\ldots h_f{}^s S^{p\ldots q}{}_{r\ldots s;b}~.
\end{equation}
This allows one --at least formally-- to remove the $n_{\alpha\beta}$ and $a_\alpha$ variables from a subset of the equations (\ref{EFE},\ref{Jacobibis},\ref{bianchi2bis},\ref{Ricci0}), 
resulting in a set consisting of 4 Jacobi equations, 4 $(0 a)$ Einstein field equations, 10 Ricci equations (\ref{Ricci0}) and the 20 Bianchi equations (\ref{bianchi2bis}) and which is usually referred to as the ``1+3 covariant equations''. 
This system is quite elegant, bears a strong analogy with the Maxwell equations\cite{MaartensBassett} (for a critique of this analogy see however \cite{CostaHerdeiro}) and forms the basis of a number of significant contributions to cosmology.
It is however an incomplete system and the fact that extra spatial information, representing
the 12 remaining Jacobi equations and 6 
Einstein field equations, must be added in one way or another, has been reported on several 
occasions\cite{Ell71, Mac_integrability, Vel97,vanElstUggla97, Uggla_etal2003}. This incompleteness problem, in combination with 
the redundancies present in the fully extended system when the missing Einstein field equations are added \emph{by hand}, was one of the main motivations for the 
present investigation. It turns out that (see section \ref{S:JRB}), when the time-like congruence corresponding to $\bm{u}$ is \emph{generic}, adding the 6 spatial field equations 
is not necessary at all. In fact we will do more and show that \emph{generically} the full set of field equations can be interpreted as 
integrability conditions for the system governed by (\ref{bianchi2}, \ref{Jacobibis}) and (\ref{Ricci0}). This system (wherein the commutator relations (\ref{commut}) 
will always be implicitly assumed to be valid) will be referred to as the Jacobi-Ricci-Bianchi (JRB) system.
The emphasis is on ``generically'', as 
the result is clearly false in certain overly simplified situations. Assuming for example a perfect fluid model 
with $\sigma_{ab}=\omega_a=\dot{u}_a=E_{ab}=H_{ab}=0$ and taking $\bm{u}$ to be the fluid's time-like eigenvector\footnote{Using the Einstein field equations, this would then 
necessarily be a  FLRW model.},
the only surviving JRB equations are 1) a single Bianchi equation (the conservation law for the matter density)
2) a single Ricci equation (the Raychaudhuri equation) and the 12 Jacobi equations for $n_{\alpha\beta}$ and $a_\alpha$, from which it is clearly impossible to derive the 
field equations.\\

In the case of a null-tetrad formalism (Newman-Penrose), the situation is less complicated, as the equations are usually already presented in their `extended' form, as sets of `NP-equations' 
and `Bianchi-equations', where the former is just the set of second Cartan equations (\ref{cartan2bis}). In section 4 I will show how also the `NP-set' can be split in
Jacobi equations and Ricci equations for one of the null congruences ($\bm{k}$), with the remaining field equations being integrability conditions of the Jacobi-Ricci-Bianchi set.

\section{The JRB-system and the Einstein field equations in a 1+3 setting}\label{S:JRB}
In order to have no misunderstanding about the meaning of the `JRB-system' I first write out the equations explicitly.
Where it is convenient, the expansion tensor $\Theta_{ab}=\sigma_{ab}+\smfrac{1}{3}\theta h_{ab}$ is used 
in stead of the shear tensor and $\theta_\alpha=\Theta_{\alpha\alpha}$. A hybrid notation will
be used, where boldface symbols refer to objects with greek indices, while
the 3D operators $\cdot$ and $\times$ have their usual meaning, 
\emph{also when dealing with non-tensorial objects}: for example $\bm{a}\cdot \bm{\omega} = a_\alpha \omega^\alpha$, $(\bmp \cdot \bm{n})_\alpha = \partial_\beta {n^\beta}_\alpha$, \ldots (see 
also \cite{VdBerghWylleman2004}).
Note that no evolution equations are obtained for the variables $\dot{\bm{u}}$ and $\bm{\Omega}$: this is not surprising as these are precisely the
frame gauge source functions\cite{Friedrich85} reflecting the freedom 
of choice of the time-like congruence $\mathbf{u}$ and of the rotation rate of the spatial triad $\partial_\alpha$. In special circumstances\cite{vanElstUggla97}, for example for a perfect 
fluid satisfying a barotropic equation of state, one may choose $\mathbf{u}$ to be the unique time-like eigenvector of the energy-momentum tensor and an evolution equation for 
$\dot{\mathbf{u}}$ is obtained from the $[\partial_0,\, \partial_\alpha]\rho$ commutator relations. A Fermi-propagated triad may be chosen to put $\mathbf{\Omega}=0$ or a co-rotating triad 
to put $\mathbf{\Omega}+\mathbf{\omega}=0$. However none of these gauge choices will be imposed in what follows.

\subsection*{Jacobi equations}\label{SS:Jac}

Writing out the 16 Jacobi equations (\ref{Jacobibis}), one obtains 12 evolution equations,

\begin{eqnarray}
\fl \partial_0 \bm{\omega} &=& -\half (\bmp -\bm{a}) \times \dot{\bm{u}}+\half \bm{n}\cdot \dot{\bm{u}}+\bm{\sigma}\cdot 
 \bm{\omega}+\bm{\omega}\times \bm{\Omega}-\smfrac{2}{3}\theta \bm{\omega}~, \label{e0_omega}\\
\fl \partial_0 \bm{a} &=&  \half \bmp\cdot \bm{\sigma}-\half (\bmp-2\bm{a}+\dot{\bm{u}})\times (\bm{\omega}+\bm{\Omega})-\smfrac{1}{3}\bmp \theta-\smfrac{1}{3}\theta (\bm{a}+
 \dot{\bm{u}}) -\bm{\sigma} \cdot (\bm{a}-\half \dot{\bm{u}})~, \label{e0_a}\\
\fl \partial_0 n_{12} &=&  \half \partial_1(\sigma_{3 1}-\omega_2-\Omega_2)-\half \partial_2(\sigma_{23}+\omega_1+\Omega_1)+\half \partial_3(\theta_2-
\theta_1) \nonumber\\
\fl &&   + (n_{1 1}+n_{22})\sigma_{1 2} + (n_{22}-n_{1 1})(\omega_3+\Omega_3) -  n_{1 2}\theta_3 \nonumber \\
\fl && +\half \dot{u}_1(\sigma_{31}- \omega_2-\Omega_2)-\half \dot{u}_2 (\sigma_{23}+\omega_1+\Omega_1) -\half \dot{u}_3(\theta_1-\theta_2) \nonumber \\
\fl &&  +  n_{3 1}(\sigma_{2 3}+\omega_1+\Omega_1)+ n_{23}(\sigma_{31}-\omega_2-\Omega_2)~, \label{e0_nab} \\
\fl \partial_0 n_{11} &=& -(\partial_1+\dot{u}_1)(\omega_1+\Omega_1)-(\partial_2+\dot{u}_2) \sigma_{31}+(\partial_3+\dot{u}_3) \sigma_{12}  \nonumber \\
\fl && +\bmp \cdot \bm{\Omega}+ 2 (\bm{a}+\dot{\bm{u}})\cdot \bm{\omega} +\bm{\Omega}\cdot \dot{\bm{u}} \nonumber \\
\fl &&+n_{11}(\theta_1-\theta_2-\theta_3) +2 n_{12}(\sigma_{12}+\omega_3+\Omega_3)+2 n_{31} (\sigma_{31}-\omega_2-\Omega_2)\label{e0_naa}
\end{eqnarray}
and 4  spatial `divergence' equations,
\begin{eqnarray}
\fl  \bmp \cdot \bm{\omega} &=& (2 \bm{a} +\dot{\bm{u}})\cdot \bm{\omega}~,\label{div_omega}\\
\fl \bmp \cdot \bm{n} &=& -\bmp \times \bm{a}+2 \bm{\Theta} \cdot \bm{\omega} + 2 \bm{n}\cdot \bm{a}+2 \bm{\omega}\times \bm{\Omega}~. \label{div_n}
\end{eqnarray}  

\subsection*{Ricci equations applied to $\bm{u}$}

Writing out (\ref{Ricci0}) and replacing, as explained above, the Riemann tensor by its decomposition in terms of 
$T_{ab}$, $E_{ab}$ and $H_{ab}$, one obtains 18 independent equations, 3 of which are just the evolution equations (\ref{e0_omega}) for the vorticity. The remaining set splits in
6  evolution equations for the expansion tensor, the trace being the Raychaudhuri equation,
\begin{equation}
 \fl  \partial_0 \theta = -\smfrac{1}{3}\theta^2- \half(\rho+3 p) - \sigma_{\alpha \beta} \sigma^{\alpha \beta} + 2 \omega_\alpha \omega^\alpha +\bmp \cdot \dot{\bm{u}} +
  \dot{\bm{u}}\cdot (\dot{\bm{u}}-2 \bm{a}) \label{Raychaudhuri} 
  \end{equation}
  and
\begin{eqnarray}
\fl \partial_0 \theta_1 &=&   \partial_1\dot{u}_1 -\smfrac{1}{9} \theta^2 -\smfrac{1}{6} (\rho+3 p )-\sigma_{11}(\sigma_{11}+\smfrac{2}{3}\theta)-\sigma_{12}^2-\sigma_{31}^2+
\omega_2^2+\omega_3^2\nonumber \\
\fl &&  +\dot{u}_1^2 +\dot{u}_2(n_{31}-a_2)-\dot{u}_3(n_{12}+a_3)+2(\sigma_{12}\Omega_3-\sigma_{31} \Omega_2) +\half \pi_{11}-E_{11}~, \label{e0_theta1}\\
\fl \partial_0 \sigma_{12} &=& \partial_{(1}\dot{u}_{2)} -\sigma_{12}(\theta_1+\theta_2)-\sigma_{31}\sigma_{23} +\sigma_{31}\Omega_1-\sigma_{23}\Omega_2+(\theta_2-\theta_1)\Omega_3 
+\dot{u}_1\dot{u}_2-\omega_1\omega_2\nonumber\\
\fl && -\half(n_{31}\dot{u}_1-n_{23}\dot{u}_2)+\half (n_{11}-n_{22})\dot{u}_3 +a_{(1}\dot{u}_{2)}+\half \pi_{12} - E_{12}~. \label{e0_sigma12}
\end{eqnarray}
There remain 9 equations for the spatial derivatives of the vorticity components, which are equivalent with the 9 `div $\bm{\sigma}$' and `curl $\bm{\sigma}$' equations of the 
1+3 covariant formalism,
\begin{eqnarray}
 \fl \partial_1 \omega_2 &=& \partial_1 \sigma_{31}-\partial_3\theta_1+(\sigma_{23}+\omega_1)(n_{31}-a_2) -2\dot{u}_1\omega_2+\half \omega_3(n_{22}+n_{33}-n_{11}) +\half q_3 +H_{12} \nonumber \\
 \fl && +\half \sigma_{12}(n_{11}+3 n_{22}-n_{33})+(n_{12}+a_3)(\theta_1-\theta_3)+2\sigma_{31}(n_{23}-a_1)~,\label{e1omega2}\\
 \fl \partial_1 \omega_3 &=& \partial_2\theta_1 -\partial_1 \sigma_{12}+(\sigma_{23}-\omega_1)(n_{12}+a_3) -2\dot{u}_1\omega_3-\half \omega_2(n_{33}+n_{22}-n_{11}) -\half q_2 +H_{31}\nonumber \\
 \fl && +\half \sigma_{31}(n_{11}+3 n_{33}-n_{22})+(n_{31}-a_2)(\theta_1-\theta_2)+2\sigma_{12}(n_{23}+a_1)~,\label{e1omega3}\\
 \fl \partial_1 \omega_1 &=& \partial_3 \sigma_{12}-\partial_2\sigma_{3 1}-\omega_1\dot{u}_1+\omega_2(\dot{u}_2+a_2-n_{3 1})+\omega_3(\dot{u}_3+a_3+n_{12}) +H_{11} \nonumber \\
 \fl && +\sigma_{3 1}(n_{3 1}+a_2)+\sigma_{12}(n_{12}-a_3) -2 n_{23}\sigma_{23}\nonumber\\
 \fl &&+\half n_{11}(3\theta_1-\theta)-\half (n_{22}-n_{33})(\theta_2-\theta_3)~. \label{e1omega1}
\end{eqnarray}
Note that with (\ref{e1omega1}), using $H_\alpha^\alpha=0$, the Jacobi equation (\ref{div_omega}) becomes an identity.

\subsection*{Bianchi equations}
Writing out the Bianchi equations (\ref{bianchi1bis}), one first obtains the 4 contracted equations, determining the evolution of $\rho$ and $q_\alpha$,
\begin{eqnarray}
 \fl \partial_0 \rho &=&  -(\rho+p) \theta - \bmp \cdot \bm{q} -2 \bm{q}\cdot (\dot{\bm{u}}-\bm{a}) -\pi_{\alpha \beta} \sigma^{\alpha\beta}~,\label{dot_rho}\\
 \fl \partial_0 q_1 &=& (-\bmp p-(\rho+p)\dot{\bm{u}}-\bmp\cdot \bm{\pi} +\bm{\pi}\cdot (3\bm{a}-\dot{\bm{u}})-\bm{\Theta}\cdot \bm{q}-\theta \bm{q})_1-q_2(\omega_3-\Omega_3)+q_3(\omega_2-\Omega_2)\nonumber \\
 \fl &&  +\pi_{31}n_{12}-\pi_{12}n_{31}-(\pi_{22}-\pi_{33})n_{23}+(n_{22}-n_{33})\pi_{23}~. \label{dot_q}
\end{eqnarray}
Then follow the sets of 10 `dot E', `dot H' and  6 `div E', `div H'  equations, the appearance of which below is somewhat more complicated than the familiar (perfect fluid) one by the presence
of the $q_\alpha$ and $\pi_{\alpha\beta}$ terms:
\begin{eqnarray}
\fl \partial_0 E_{12} &=& -\half{\it \partial}_{{2}} q_{{1}}  -\half\,{\it \partial}_{
{0}} \pi_{{12}}  +{\it \partial}_{{2}}  H_{{23}} -{\it \partial}_{{3}}  H_{{22}} -\half \left(\omega_{{3}}+\sigma_{{12}} \right) (\rho+p) \nonumber \\
\fl && + \smfrac{1}{4}\left(n_{{22}
}-n_{{33}}-n_{{11}} \right) q_{{3}}+ \half\left( -n_{{23}}-a_{{1}}-\dot{u}_{{1}} \right) q_{{2}}-\half \dot{u}_{{2}}q_{{1}}\nonumber \\
\fl && -\left( \Omega_{{2}}-2\,\sigma_{{31}}+2\,\omega_{{2}} \right) E_{{23}
}+ \left( \sigma_{{23}}+\Omega_{{1}}-\omega_{{1}} \right) E_{{31}}- \left( \theta_{{2}}+2\,\theta_{{3}} \right) E_{{12}}\nonumber \\
 \fl && + 
(2 \Omega_{{3}}+\sigma_{12}+\omega_3)E_{{22}}+\left(\Omega_3- \sigma_{{12}}-\omega_{{3}} \right) E_{{33}} \nonumber \\
\fl &&  + \left( \dot{u}_{{2}}-2\,a_{{2}}-2\,n_{{31}} \right) H_{{23}}- \half \left( 3 n_{{11}}+n_{{22}}-n_{{33}} \right) H_{{12}} 
- \left( a_{{1}}+\dot{u}_{{1}}+n_{{23}} \right) 
H_{{31}}\nonumber \\
\fl && + \left( a_{{3}}-n_{{12}}-2 \dot{u}_{{3}} \right) H_{{22}} + \left( n_{{12}}-a_3-\dot{u}_3 \right) H_{{33}}  \nonumber \\
\fl &&  -\half(
(\sigma_{{12}}+\omega_{{3}}+\Omega_{{3}})\pi_{{11}}+\Omega_{{2}}\pi_{{23}}+\theta_
{{2}}\pi_{{12}}-\Omega_{{3}}\pi_{{22}}-\left( \Omega_{{1}}-\sigma_{{
23}}+\omega_{{1}} \right) \pi_{{31}})~, \label{dot_E12}\\
\fl \partial_0 E_{11} &=& -\half {\it \partial}_{{0}} \pi_{{11}} -\smfrac{1}{2}{\it \partial
}_{{1}} q_{{1}} +\smfrac{1}{6} \bmp \cdot \bm{q} 
 +{\it \partial}_{{2}}  H_{{31}} -{\it \partial}_{{3}} H_{{12}} +\smfrac{1}{6}\left(\theta-3\theta_{{1}} \right) (\rho+p)
\nonumber \\
\fl &&
 -\smfrac{1}{3}\left( 2\dot{u}_{{1}}+a_{{1}} \right) q_{{1}}+ \smfrac{1}{6}\left(a_{{2}
}+2 \dot{u}_{{2}}-3 n_{{31}} \right) q_{{2}}+ \smfrac{1}{6}\left( a_{{3}}+3 n_{{12}}+2\dot{u}_{{3}} \right) q_{{3}}
\nonumber \\
\fl && +
 \left( \sigma_{{12}}+\omega_{{3}}+2\,\Omega_{{3}} \right) E_{{12}}-2
\,E_{{23}}\sigma_{{23}}+ \left( \sigma_{{31}}-\omega_{{2}}-2\,\Omega_{
{2}} \right) E_{{31}}
\nonumber \\
\fl &&
-\left( \theta_{{2}}+2\,\theta_{{3}} \right) E_
{{11}}+ \left( \theta_{{2}}-\theta_{{3}} \right) E_{{33}}
-\smfrac{3}{2} H_{{11}}n_{{11}}+ \half \left(n_{{22}}-n_{{33}}
 \right) (H_{{22}}-H_{{33}})\nonumber \\
 \fl && +2\,H_{{23}}n_{{23}}-\left( n_{{31}}-2\,\dot{u}_{{2}}+a_{
{2}} \right) H_{{31}}-\left( n_{{12}}+2\,\dot{u}_{{3}}-a_{{3}} \right) H_{
{12}}
\nonumber \\
\fl &&  -\smfrac{1}{6}\left(\sigma_{{31}}+ 3\omega_{{2}}+6\Omega_{{2}} \right) \pi
_{{31}}-\smfrac{1}{6} \left( \sigma_{{12}}-3\omega_{{3}}-6 \Omega_{{3}}
 \right) \pi_{{12}}+\smfrac{1}{3}\pi_{{23}}\sigma_{{23}} \nonumber \\
 \fl && - \smfrac{1}{6}\left( 2\theta_{{1}}+\theta_{
{3}} \right) \pi_{{11}}+ \smfrac{1}{6}\left(\theta_{{2}}-\theta_{{3}}
 \right) \pi_{{22}}~, \label{dot_E11}
\end{eqnarray}
\begin{eqnarray}
\fl \partial_0 H_{12} &=& \partial_3(E_{22}+\smfrac{1}{6}\rho+\half \pi_{11})-\half{\it \partial}_{{1}} \pi_{{31}} -{\it \partial}_{{2
}} E_{{23}} + \half \left(\sigma_{{31}}-3\omega_{{2}} \right) q_{{1}}-\half q_{{3}}\theta_{{1}} \nonumber \\
\fl && + \half \left( 3 n_{{11}}+n_{{22}}-n_{{33}} \right) E_{{12}}+ \left(n_{{23}}+a_{{1}}+\dot{u}_{{1}} \right) E_{{31}}+
 \left( 2\,n_{{31}}-\dot{u}_{{2}}+2\,a_{{2}} \right) E_{{23}} \nonumber \\
\fl && + \left( 2\,a_{
{3}}-2 n_{12}-\dot{u}_{{3}} \right) E_{{33}}- \left(n_{{12}}+2\,\dot{u}_{{3}}-a_{{3}} \right) E_{{11}} \nonumber \\
\fl &&  +\left( 2\,\sigma_{{31}}-2\,\omega_{{2}} -\Omega_{{2}}\right) H_{{23}
}+ \left( \sigma_{{23}}-\omega_{{1}}+\Omega_{{1}} \right) H_{{31}}
- \left( \theta_{{2}}+2\,\theta_{{3}} \right) H_{{12}} \nonumber \\
\fl && +(H_{22}-H_{{11}})\Omega_{{3}}+ \left( \sigma_{{12}}+\omega_{{3}}
 \right)( H_{{22}}-H_{33})-\half \left( a_{{3}}+n_{{12}} \right) (\pi_{11}-\pi_{{33}} ) \nonumber \\
\fl && +\smfrac{1}{4}\left(n_{{33}}-n_{{11}}-3 n_{{22}} \right) \pi_{{12}}+
 \left( a_{{1}}-n_{{23}} \right) \pi_{{31}}+ \half \left(a_{{2}}-n_{{31}} \right) \pi_{{23}}~, \label{dot_H12}\\
 \fl \partial_0 H_{11} &=& {\it \partial}_{{3}}  (E_{{12}}-\half \pi_{12}) -{\it \partial}_{{2}} ( E_{{31}} -\half \pi_{{31}})
 -q_{{1}}\omega_{{1}}+ \half \left(\omega_{{2}}-\sigma_{{31}}
 \right) q_{{2}}+ \half \left(\sigma_{{12}}+\omega_{{3}} \right) 
q_{{3}}
\nonumber \\
\fl &&  + \left( n_{{12}}-a_{{3}}+2\,\dot{u}_{{3}} \right) E_{{12}}-2\,E_{{23}}n_{{23
}}+ \left( n_{{31}}+a_{{2}}-2\,\dot{u}_{{2}} \right) E_{{31}} \nonumber \\
\fl && + \half \left( n_{{33}}-3 n_{{11}}-n_{{22}} \right)( E_{{22}}-\half \pi_{22})+ \half
 \left(n_{{22}} -3 n_{{11}}-n_{{33}} \right) (E_{{33}}-\half \pi_{33}) \nonumber \\
\fl &&+ H_{{22}} (\theta_2+2 \theta_3) +H_{{33}}(\theta_3+2 \theta_2) + \left( \sigma_{{12}}+\omega_{{3}}+2\,\Omega_{{3}} \right) H_{{12}}+ \left( \sigma_{{31}}-\omega_{{2}}-2\,\Omega_{
{2}} \right) H_{{31}}\nonumber \\
\fl &&  -2\,\sigma_{{23}}H_{{23}}+\half \left( n_{{22}}-n_{{33}} \right) \pi_{{22}}+n_{{23}} \pi_{{23}} -\half \left( n_{{12}}-a_{{3}} \right) \pi_{{12}} -
\half \left(n_{{31}} +a_2\right) \pi_{{31}} \label{dot_H11}
\end{eqnarray}
and
\begin{eqnarray}
 \fl (\bmp \cdot \bm{E})_3 &=& (\smfrac{1}{3} \bmp \rho -\half \bmp \cdot \bm{\pi} + \smfrac{3}{2} \bm{\omega}\times \bm{q} +\half \bm{\Theta}\cdot \bm{q}+ 3\bm{E}\cdot \bm{a}-3\bm{H}\cdot \bm{\omega} +\smfrac{3}{2}\bm{\pi}\cdot \bm{a})_3
\nonumber \\
\fl && + n_{{31}} E_{{23}} - n_{{23}} E_{{31}} +
 \left( n_{{11}}-n_{{22}} \right) E_{{12}}  + (E_{{22}}  - E_{{11}}) n_{{12}}  
\nonumber \\
\fl &&  + \sigma_{{12}} (H_{22}-H_{11}) + \left( \theta_{{1}}-\theta_{{2}} \right) H_{{12}}+ \sigma_{{31}}H_{{23}} - \sigma_{{23}} H_{{31}}  -\half n_{{23}}\pi_{{31}}+ \half n_{{31}} \pi_{{23}}
\nonumber \\
\fl && +\half (n_{{12}}-3 a_3)\pi_{22}-\half (n_{{12}}+3 a_3)\pi_{11} + \half \left(n_{{11}}-n_{{22}} \right) \pi_{{12}}~, \label{div_E}\\ 
  \fl (\bmp \cdot \bm{H})_3  &=& (-\half \bmp \times \bm{q}+  (\rho +p)\bm{\omega} + \half \bm{a}\times \bm{q} +\half \bm{n}\cdot \bm{q}+3 \bm{E}\cdot \bm{\omega}+3 \bm{H}\cdot \bm{a}-\half  \bm{\pi}\cdot \bm{\omega})_3
\nonumber \\
\fl &&  + \left( \theta_{{2}}-\theta_{{1}} \right) E_{{12}}- \sigma_{{
31}} E_{{23}}+ \sigma_{{23}} E_{{31}} +\sigma_{12}(E_{11} -E_{22})  \nonumber \\
\fl && + n_{{31}} H_{{23}} - n_{{23}} H_{{31}} + \left( n_{{11}}-n_{{22}} \right) H_{{12}} +n_{{12}} (H_{{22}}-H_{11}) 
\nonumber \\
\fl && - \half \left( \theta_{{1}}-\theta_{{2}} \right) \pi_{{12}}
+\half  \sigma_{{23}} \pi_{{31}}
-\half  \sigma_{{31}} \pi_{{23}}
+\half  \sigma_{{12}}( \pi_{{11}}- \pi_{{22}} )\label{div_H}~.
\end{eqnarray}

To sum up: the JRB-system consists of the following independent sets: 15 Jacobi equations (12 evolution equations (\ref{e0_omega}--\ref{e0_naa}) and 3 `div $\bm{n}$' equations (\ref{div_n})), 
15 Ricci equations (6 evolution equations
(\ref{Raychaudhuri}--\ref{e0_sigma12}) and 9 equations (\ref{e1omega2}--\ref{e1omega1}) for the spatial derivatives of the vorticity) and 20 Bianchi equations 
(14 evolution equations (\ref{dot_rho},\ref{dot_q},\ref{dot_E12}--\ref{dot_H11}) and 6 `div $\bm{E}$', `div $\bm{H}$' equations (\ref{div_E},\ref{div_H})). 

\subsection*{Einstein field equations}
We now write out the Einstein field equations, simplifying the left hand side of (\ref{EFE}) by using all the algebraic information in the 
Jacobi and Ricci equations. The $(0\alpha)$ components become then identically satisfied, while  the $(00)$ equation 
reduces to the trace of the $(\alpha\alpha)$ field equations. We write the remaining 6 equations as $\FE_{\alpha\beta}=0$, 
with
\begin{eqnarray}
 \fl \FE_{11} &\equiv&  {\it \partial}_{{2}} (n_{{31}}+a_2)-{\it \partial}_{{3}}( n_{{12}}-a_3)+ \theta_{{2}}\theta_{{3}}-{\sigma_{{23}}}^{2}+{\omega_{{1}}}(\omega_1-2\Omega_1) -a_\alpha a^\alpha \nonumber \\
\fl && +\smfrac{1}{4}n_{11}(2 n_{{22}}+2 n_{{33}} -3n_{11})+\smfrac{1}{4}(n_{22}-n_{33})^{2}+{n_{{23}}}^{2} \nonumber \\
\fl && -{n_{{12}}}(n_{12}-2 a_{{3}}) -{n_{{31}}}(n_{31}+2 a_{{2}})+E_{{11}}-\smfrac{1}{3}\rho+\half \pi_{{11}} \label{efe11} 
\end{eqnarray}
and
\begin{eqnarray}
\fl \FE_{12} &\equiv& \half {\it \partial}_{{1}}(n_{31}+ a_{{2}} ) -\half {\it \partial}_{{2}} (n_{{23}} - a_{{1}})
-\half{\it \partial}_{{3}}  (n_{{11}} -  n_{{22}}) +\omega_{{1}}\Omega_{{2}}+\omega_{{2}}
\Omega_{{1}}-\omega_{{2}}\omega_{{1}}\nonumber \\
\fl && -\sigma_{{31}}\sigma_{{23}}+\theta_{{3}}\sigma_{{12}}+ n_{{12}}(n_{{11}}+n_{{22}}-n_{{33}})  +2\,n_{{31}}n_{{23}} \nonumber \\
\fl && -n_{{31}}a_{{1}}+n_{{23}}a_{{2}}  +a_{{3}}(n_{{11}}-n_{{22}} ) -E_{{12}}-\half \pi_{{12}}~.
\end{eqnarray}
It might appear odd that the electric part $E_{ab}$ of the 
Weyl tensor shows up in the left hand side of (\ref{EFE}), but this is of course a consequence of the fact that we have eliminated the evolution of the shear in favour of $E_{ab}$ by means of the 
Ricci equations (\ref{e0_theta1},\ref{e0_sigma12}).

\subsection*{Integrability conditions for the JRB-system}
The JRB-system provides us with expressions for all derivatives of the vorticity, in terms of
spatial derivatives of the expansion tensor and the acceleration. It is relatively easy to verify that the integrability conditions for these equations are identically satisfied under the
Einstein field equations (see also \cite{Els96}, where the propagation of the constraints 
under the evolution equations was demonstrated for a barotropic perfect fluid, or \cite{Uggla_etal2003} for a general energy-momentum tensor and a non-rotating congruence). That the 
field equations can also be viewed as integrabily 
conditions for the JRB-system (provided the $\bm{u}$ congruence is generic) is not evident at all, but can be seen by evaluating the following set of commutators, 
\begin{eqnarray}
 &[\partial_0, \partial_1](\sigma_{31}+\omega_2)+[\partial_0, \partial_3]\theta_1+[\partial_3, \partial_1]\dot{u}_1~,\\
 & [\partial_0, \partial_1](\sigma_{12}+\omega_3)-[\partial_0, \partial_2]\theta_1+[\partial_1, \partial_2]\dot{u}_1~, \\
 & [\partial_1, \partial_2](\sigma_{23}+\omega_1)+[\partial_2, \partial_3](\sigma_{12}-\omega_3)+[\partial_3,\, \partial_1 ] \theta_2.
\end{eqnarray}
From the resulting expressions (and from similar ones obtained by cyclic permutation of the indices) all second order derivatives can be eliminated, leading
to a homogeneous system of first order equations with the following simple structure:
 \begin{eqnarray}
 \left[ \begin{array}{lll}
       \dot{u}_3 & 0 &-\dot{u}_2  \\
       -\dot{u}_3 & \dot{u}_1 & 0 \\
       0 &-\dot{u}_1 &\dot{u}_2
      \end{array}\right]  \ \,  \left[
      \begin{array}{c}
       \FE_{12}\\ \FE_{23} \\ \FE_{31}
      \end{array}\right] = 0~, \label{matrix1} \\
 \textrm{diag}(\dot{u}_1,\dot{u}_2,\dot{u}_3)  \left[ \begin{array}{c}
       \FE_{33}\\ \FE_{11} \\ \FE_{22}
      \end{array}\right] -   \left[
   \begin{array}{lll}
    0 &0 &\dot{u}_3\\
    \dot{u}_1&0&0 \\
    0&\dot{u}_2 &0 
   \end{array}\right] \ \,  \left[ \begin{array}{c}
       \FE_{12}\\ \FE_{23} \\ \FE_{31}
      \end{array}\right]
   = 0~, \label{matrix2}\\
   \textrm{diag}(\dot{u}_2,\dot{u}_3,\dot{u}_1)  \left[ \begin{array}{c}
       \FE_{33}\\ \FE_{11} \\ \FE_{22}
      \end{array}\right] -   \left[
   \begin{array}{lll}
    0&-\dot{u}_3  &0\\ 
    0 &0 &\-\dot{u}_1 \\
    -\dot{u}_2&0&0 
   \end{array}\right] \ \,  \left[ \begin{array}{c}
       \FE_{12}\\ \FE_{23} \\ \FE_{31}
      \end{array}\right]
   = 0~, \label{matrix3}\\
 \left[  \begin{array}{lll}
    \sigma_{23}+\omega_1 &0& -\sigma_{23}+\omega_1 \\
    0&-\sigma_{12}+\omega_3&\sigma_{12}+\omega_3 \\
    -\sigma_{31}+\omega_2 &\sigma_{31}+\omega_2& 0
   \end{array}\right] \ \,  \left[
   \begin{array}{c}
       \FE_{33}\\ \FE_{11} \\ \FE_{22}
      \end{array}\right]  \nonumber \\ 
 \lo+ \left[
      \begin{array}{lll}
       -\sigma_{31}-\omega_2&\theta_2-\theta_3 &\sigma_{12}-\omega_3\\
       \theta_1-\theta_2&\sigma_{31}-\omega_2& -\sigma_{23}-\omega_1 \\
       \sigma_{23}-\omega_1& -\sigma_{12}-\omega_3& \theta_3-\theta_1 
      \end{array} \right] \ \,  \left[
      \begin{array}{c}
       \FE_{12}\\ \FE_{23} \\ \FE_{31}
      \end{array} \right] = 0~.\label{matrix4}
 \end{eqnarray}
Aligning the $\partial_3$ axis with the acceleration ( $\dot{u}_1=\dot{u}_2=0$), it is clear that, for 
non-vanishing acceleration, $\FE_{12}=\FE_{23}=\FE_{31}=\FE_{11}=\FE_{22}=0$,
 which when substituted in (\ref{matrix4}) implies $(\sigma_{31}-\omega_2)\FE_{33}=(\sigma_{23}+\omega_1)\FE_{33}=0$. It follows that $\FE_{33}=0$ unless 
 $\sigma_{\alpha 3}+\omega_{\alpha 3}=0$, meaning that 
 $\dot{\bm{u}}$ is an eigenvector of $\sigma_{ab} +\omega_{ab}$:
 \begin{thm}[1]
  If $\dot{\bm{u}}\neq 0$ is not an eigenvector of $\sigma_{ab} +\omega_{ab}$, then the Einstein
  field equations are the integrability conditions of the JRB-system.
 \end{thm}
 
 Notice that this theorem is also an immediate consequence of the property that the curvature tensor for 
 a given metric connection $\nabla$ is the \emph{unique} tensor obeying 
 \begin{itemize}
  \item the symmetries (\ref{Riesymm})
  \item the first Bianchi equations (\ref{bianchi1bis})
  \item the second Bianchi equations (\ref{bianchi2bis})
  \item the Ricci equations (\ref{Ricci0}) for a generic $\bm{u}$ congruence.
 \end{itemize}
 A simple covariant proof of this uniqueness property is obtained by introducing the difference $\bm{A}$ of two such tensors and contracting (\ref{bianchi2bis}) with $u^b$, implying
 \begin{equation}\label{main1}
  A_{ab[cd} {u^b}_{;e]}=0~.
 \end{equation}
By (\ref{Ricci0}) one has $A_{abcd}u^d=0$ and hence a further contraction of (\ref{main1}) with $u^e$ implies $A_{abcd}\dot{u}^d=0$. Defining a tetrad with $\partial_0=\bm{u}$ and 
$\partial_3=\dot{\bm{u}}$,
it follows that the only possible non-zero component of $\bm{A}$ is given by $A_{1212}$ ($=A_{2121}=-A_{1221}=-A_{2112}$). Choosing in (\ref{main1}) $(acde)=(1123)$ or $(2213)$ gives then 
\begin{equation}
 A_{2121}{u^1}_{;3} = A_{1212}{u^2}_{;3} = 0,
\end{equation}
which indeed implies $\bm{A}=0$, unless ${\dot{u}}_{[c} u_{a];b}\dot{u}^b=0$, i.e.~unless $\dot{\bm{u}}$ is an eigenvector of $\sigma_{ab}+\omega_{ab}$.
 
A similar property ---not involving the acceleration of the time-like congruence--- can be obtained by considering the integrability conditions of the second Bianchi equations:
eliminating the second order derivatives from the commutator relations
$[\partial_0,\, \partial_1] H_{11} + [\partial_0,\, \partial_2] H_{12} + [\partial_0,\, \partial_3] H_{13}-\smfrac{1}{6}[\partial_2,\, \partial_3] \rho
- [\partial_3,\,   \partial_1]E_{12}-[\partial_1,\, \partial_2] E_{31} -[\partial_2,\, \partial_3] E_{11}$ and $
[\partial_0,\, \partial_1] E_{11} + [\partial_0,\, \partial_2] E_{12} + [\partial_0,\, \partial_3] E_{13}-\smfrac{1}{3}[\partial_0,\, \partial_3] \rho
+ [\partial_3,\,   \partial_1]H_{23}+[\partial_1,\, \partial_2] H_{33} +[\partial_2,\, \partial_3] H_{31}$
again results in a homogeneous system for the $\FE_{\alpha\beta}$ variables:
 \begin{equation}
 \fl  \left[ \begin{array}{lll}
       -\E_{23}&\hfill 0&\hfill \E_{23}\\ \hfill\E_{31}&-\E_{31}& \hfill 0\\ \hfill 0& \hfill \E_{12}&-\E_{12}
      \end{array}\right]   \,  \left[
      \begin{array}{c}
      \FE_{33}\\ \FE_{11} \\ \FE_{22}
      \end{array}\right] =  \left[
      \begin{array}{lll}
      \hfill -\E_{31}&\E_{22}-\E_{33}&\hfill \E_{12}\\
      \hfill \E_{23}&\hfill -\E_{12}&\E_{33}-\E_{11}\\
      \E_{11}-\E_{22}&\hfill \E_{31}&\hfill -\E_{23}
      \end{array}  \right] \, 
            \left[ \begin{array}{c}
       \FE_{12}\\ \FE_{23} \\ \FE_{31}
      \end{array}\right]~, \label{matrix5}
 \end{equation}
 \begin{equation}
 \fl  \left[\begin{array}{lll}
       \hfill 0&-\H_{21}&\hfill \H_{12}\\-\H_{13}&\hfill \H_{31}&\hfill 0\\ \hfill \H_{23}&\hfill 0&-\H_{32}
      \end{array}\right]   \,  \left[
      \begin{array}{c}
      \FE_{33}\\ \FE_{11} \\ \FE_{22}
      \end{array}\right] = \left[ 
      \begin{array}{lll}
      \H_{22}-\H_{11}&\hfill -\H_{13}&\hfill \H_{23}\\
      \hfill -\H_{32}&\hfill \H_{12}&\H_{11}-\H_{33}\\
      \hfill \H_{31}&\H_{33}-\H_{22}&\hfill -\H_{21}
      \end{array} \right] \, 
            \left[ \begin{array}{c}
       \FE_{12}\\ \FE_{23} \\ \FE_{31}
      \end{array}\right]  \label{matrix6}    
 \end{equation}
with $\E_{ab}=E_{ab}-\smfrac{1}{2}\pi_{ab}$ and $\H_{ab}=H_{ab}-\smfrac{1}{2}\eta_{abcd} q^c u^d$.

Generically the corresponding 6x6 determinant is different from zero and the system will again have only the zero-solution. Except for a few particular cases it is not easy however to provide a 
simple geometric characterisation of the non-vanishing of this determinant. It is clear that the left hand side of (\ref{matrix5}) is zero in an $\bm{\E}$ eigenframe, such that $\FE_{12}=\FE_{23}=\FE_{31}=0$ when 
$\bm{\E}$ is non-degenerate. If $\bm{\E}$ and $\bm{H}$ commute a common eigen-frame can therefore be constructed in which clearly also $\FE_{11}=\FE_{22}=\FE_{33}=0$ provided 
$q_1q_2q_3 \neq 0$ (this will happen for example when $\mathbf{u}$ is the time-like eigenvector of a perfect fluid energy-momentum tensor for which $\mathbf{E}$ and $\mathbf{H}$ commute and hence 
a fortiori occurs for all purely electric or purely magnetic perfect fluids).  Hence we obtain
\begin{thm}[2]
  When $[\bm{\E}, \bm{H}]=0$, $\E_{\alpha\beta}$ is not degenerate and $\bm{q}$ is not parallel to one of the eigenblades of $\bm{H}$, then the Einstein
  field equations are the integrability conditions of the JRB-system.
 \end{thm}
 On the other hand it is easy to verify that the Einstein field equations \emph{cannot} be obtained from (\ref{matrix5},\ref{matrix6}) when for example $\bm{q}=0$, as then the matrix in the left hand side 
 of (\ref{matrix6}) is singular.
 
\section{The case of a null congruence}

When the congruence, with respect to which the Ricci equations are constructed, is null, one can construct a Newman-Penrose tetrad $(\bm{m}, 
\overline{\bm{m}},\bm{\ell},\bm{k})$ with $\bm{k}=\bm{u}$: the equations are then usually presented\cite{PenroseRindler,Kramer} in their `already extended' form, as the set of `NP-equations' ---these
being just the second Cartan equations (\ref{cartan2bis})--- and the second Bianchi equations. We will refer to the two sets by their numbering in \cite{Kramer} as $NP_1$--$NP_{18}$ and 
$B_1$--$B_{11}$, with the latter being just the set of integrability conditions of the former. One may again isolate from the NP-equations the subsets of 
Jacobi and Ricci equations and ask whether the remaining part of the NP-equations (the `remaining' field equations) is obtainable as integrability conditions of the 
Jacobi-Ricci-Bianchi set. An explicit evaluation of (\ref{Jacobi}) shows that the Jacobi-part of the NP-equations is expressed by the 16 independent equations (4 imaginary and 6 complex)
\begin{eqnarray}\label{JacNP}
\fl && \overline{NP_1}-NP_1,\,\overline{NP_{14}}-NP_{14},\, \overline{NP_{12}}+\overline{NP_8}-NP_{12}-NP_8, \nonumber \\
&& \overline{NP_6}-NP_6+NP_8-\overline{NP_8}, \, NP_{16}-\overline{NP_7},\,NP_{17}+NP_{8}, \nonumber \\
 && 2 NP_4 -\overline{NP_3}-\overline{NP_{11}},\, 2 NP_5 -NP_3+NP_{11}, \nonumber \\
 && 2 NP_{15} -\overline{NP_9}-\overline{NP_{13}},\,  2 NP_{18}+NP_9-NP_{13}~.
\end{eqnarray}
The Ricci equations (\ref{Ricciid}) for $\bm{k}$ on the other hand give rise to 18 independent equations (1 real, 1 imaginary and 8 complex)
\begin{eqnarray}\label{RicNP}
 && NP_1,\, NP_2,\, NP_3,\, NP_{11},\, NP_{16},\, NP_{17}, \, \overline{NP_4}+NP_5, \nonumber \\
 && NP_{15}-\overline{NP_{18}},\, \overline{NP_{6}}+NP_{6}, \overline{NP_{12}}-NP_{12}~,
\end{eqnarray}
which together with (\ref{JacNP}) reduce the remaining field equations to a set of 6 equations (2 real and 2 complex),
\begin{equation}\label{FEQNP}
 NP_{10},\, NP_{13}, \, NP_{12}+\overline{NP_{12}},\, NP_{14}+\overline{NP_{14}}
\end{equation}
(or any set equivalent with it under (\ref{JacNP},\ref{RicNP})).
Considering the linear combinations of commutators $[\overline{\delta},\, \delta] \kappa +[\delta, \, D ] \rho-[\overline{\delta},\, D] \sigma$, $[\Delta, \, D]\rho+[\overline{\delta},\, \Delta]\kappa
-[\overline{\delta},D] \tau $, $[\Delta , \, D] \alpha-[\overline{\delta},\, D]\gamma +[\overline{\delta},\, \Delta] \epsilon$ and $[\Delta , \, D] \beta-[\delta,\, D]\gamma +[\delta,\, \Delta] \epsilon$ one 
obtains then, using (\ref{JacNP},\ref{RicNP}), the equations $\kappa NP_{13}$, $\kappa NP_{12}$, $\overline{\kappa} NP_{14}-\kappa NP_{10}$ and 
\begin{equation*}
 D(NP_{13}) +\overline{\sigma} \overline{NP_{13}}-(\rho -2 \epsilon) {NP_{13}}-2(\pi+\tau) NP_{12}+2 \kappa NP_{10}~,
\end{equation*}
from which, provided $\kappa \neq 0$, (\ref{FEQNP}) readily follows:
\begin{thm}[3]
  If $\kappa=-k_{a;b} m^a k^b \neq 0$ the Newman-Penrose equations $NP_{10}, NP_{13}$, $NP_{12}+\overline{NP_{12}}$ and $NP_{14}+\overline{NP_{14}}$ are the integrability conditions of the
  remaining Newman-Penrose equations and Bianchi equations.
 \end{thm}
 
\section{Conclusion}
Embedding the 1+3 covariant equations in an extended tetrad formalism\cite{EMM} leads to redundancies, forcing one to make a choice among the sets of Jacobi, Ricci, Bianchi equations and
Einstein field equations (whereby with `Ricci equations' we mean the full set of Ricci equations applied to the tangent vectorfield of a \emph{single} time-like congruence). It is shown that
a minimal set of equations can consist of 15 Jacobi and 15 Ricci equations, together with the 20 Bianchi equations, in which the Riemann tensor is defined in terms of trace-free and symmetric
tensors $E_{ab}, H_{ab}$ and a symmetric tensor $T_{ab}$. The Einstein field equations arise then as the integrability conditions for this set, if one
chooses the time-like congruence such that the acceleration is not zero and is not an eigenvector of $\sigma_{ab}+\omega_{ab}$, a condition which is obviously violated when
 $\boldsymbol{u}$ is the 4-velocity for pressure-free matter. It remains to be seen whether this alternative view of the Einstein equations, as integrability conditions of an extended system, has any useful applications at all, for example in the area of numerical relativity. In the case of a null congruence 
$\bm{u}=\bm{k}$ an analoguous result is obtained provided that the associated Newman-Penrose coefficient $\kappa=-k_{a;b} m^a k^b$ is non-vanishing.

\section*{Acknowledgement}
I thank Stanley Deser, Con Lozanovski and Lode Wylleman for comments and suggestions for improvement. All calculations were done with the aid of the Maple symbolic algebra package.

\end{document}